# Semantics and pragmatics in actual software applications and in web search engines: exploring innovations.

Fabrizio M.A. Lolli

*Università Guglielmo Marconi, Roma, Italy, fabrizio@lolli.us*

**Abstract:** While new ways to use the Semantic Web are developed every week, which allow the user to find information on web more accurately - for example in search engines - some sophisticated pragmatic tools are becoming more important - for example in web interfaces known as *Social Intelligence*, or in the most famous Siri by Apple.

The work aims to analyze whether and where we can identify the boundary between semantics and pragmatics in the software used by analyzed systems. examining how the linguistic disciplines are fundamental in their progress. Is it possible to assume that the tools of social intelligence have a pragmatic approach to the questions of the user, or it is just a use of a very rich vocabulary, with the use of semantic tools?

## Introduction

Charles Morris in 1938 defined[1] semiotics built of three components: syntax, semantics, and pragmatics. These words were used by the computer scientists. Most part of web users see as a symbolic system. It was born with this character, but, in the last years, semantic web is become important. Even if semantic web has become more serious, one can not ignore the basic syntactic web, and the future is tuned to a pragmatic approach. To follow the path that seems to lead to a pragmatic web, we analyze the evolution of the most well-known search engine, Google, and one of the applications that, potentially at least, appears to be the turning point pragmatics of communication via the web: Apple's Siri.

---

[1] Charles Morris, editor. Foundations of the Theory of Signs. University of Chicago Press, Chicago and London, 1938.

*«When you ask a question, you do not want to get 10 million answers. You want the 10 best answers.»[2]*

## What Is Semantic Search?

Planning a search with a search engine like Google, it comes a list of results based on the text digitized. In few words, syntactic search is based on written words. The results will be the same independently from the person who type the text in the search query box.

Semantic search uses a sort of artificial intelligence (AI) in order to understand the searcher's intent and the meaning of the query, even if the data base is still the same, a dictionary, this kind of search uses something from your personal data, your character, your life style, in order to give you the best answer to your question. In a Semantic Search, Google will surf from word to word, how they are related each other, and how they could be used in your search.

Here's an example: in a restaurant, we ask for the menu. All what is on the menu is eatable? There's soup and meal in the list, but also wine, beer, and maybe service's price. There are big or small portions, but also a phone number and an address of the restaurant. All these things are not eatable!

Semantic on the web is not a recent concept: from 2008, all majors search engines began to research on natural language keywords.

Natural language understanding - NLU - is a subtopic of natural language processing in artificial intelligence that deals with machine reading comprehension. Dan Miller, senior analyst and founder, Opus Research, says «NLU should allow people to speak naturally and have a reasonable expectation that a machine on the other end is going to understand their intent.»[3]

It is true that the technology behind the creation of the software for voice recognition is actually geared towards the recognition of the intent of the user, or of what the user would like to ask the machine? Or is it a truly rapid use of an extremely large database, simply compared to the sequence of letters typed by the user or pronounced?

---

[2] Bruno Henry, Wired, nov.1999 # 7, page 99.

[3] Michele Masterson, "Natural Language Understanding Grows Up: The Bar Has Been Raised for Technologies That Not Only Hear, but Understand Too" CRM Magazine, May 2012.

The first definition, often cited, is that of Tim Berners-Lee, which, in 2001, proposed the term Semantic Web for the first time: «The Semantic Web is an extension of the current Web in which the information has a highly definite meaning and in which computers and users work in cooperation.»[4]

The term has been associated, from that moment, an idea of the web in which agents act - defined by some *intelligent agents* - able to understand the meaning of the texts present on the network, linked by paths articulated thanks to the information contained in them, metadata.

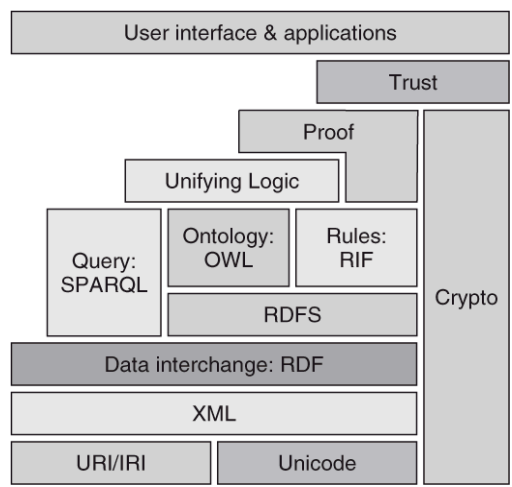

Figure 1. Semantic Web technology stack.

*Source: www.w3.org/DesignIssues/diagrams/sweb-stack/2006a.png*

The agents, active part of the software and regulated by the search algorithms must be able to direct the user to the information he seeks. Moreover, they should be able to replace it in some operations, starting from the most common and recurring (e.g. "What's the weather doing? What time is it?").

---

[4] Berners-Lee, Tim, Hendler, James, & Lassila, Ora. (2001, May 17). The semantic web. Scientific American.com.

Agents must be able to follow the directions of metadata, which connect all the information available to the user, and searched independently, to reply to the user.

Actually the actual language used and RDF, W3C is that as the correct language for the semantic web, it seems to represent a new approach to the problem (Figure 1).

The use of *triple* is the basis of this language: subject, predicate, value. With this system you can potentially connect any meaning in the web.

The original question, however, is figuring out how to work the systems underlying semantic web technology, but whether this definition and the next one in the list (NIELSEN 1996), the pragmatic web is actually assimilated or at least partially adopted by current tools of research and use of the web.

## Apple Siri, Google Voice Search, IBM Watson

The Google Voice Search is a tool that implements the Google search on the web, allowing the user to ask voice, directly to your computer, the information required. It was also considered the Apple system called Siri, which is used from October 2011 on iOS devices, and called "voice assistant".

Everything to do with the actual operation of the software of Apple's Siri, developed by Nuance and then bought by Apple in 2010, as well as all the information concerning him, are not public. Requests addressed to different companies and structures involved in the process of development and updating of the interface do not respond or respond in a negative way. Then we try to understand the functioning of the interface from the available literature and from direct experience.

Despite the public for the use of the interface Apple's Siri or Google Voice Search may seem very similar, only in the first case we have an attempt to semantic search, the system called NLU (Natural Language Understanding), which analyzes the phrase or parts of it uttered by the user.

Roberto Pieraccini, CEO of ICSI, explains: «In the case of words that are spoken into a text box (i.e. Google Voice Search), it does not mean that the machine understood what you said, but can translate sounds into words. But if you go to Siri, the meaning of the words can be understood, because there is a second part of the application, which is natural language understanding.»

So the NLU seems to be the next step and prerequisite for the manifestation of a real semantic web approach. Return again for a moment to clarify what is meant by semantic web: a part of the web where user information to be understood and interpreted, to provide a correct response to the needs expressed. One of the tools that can be considered at a great level, at this time, it is called Watson, and is a system developed by IBM in 2011. Watson took part in the TV show *Jeopardy!*. During the tv show three contestants have to choose topics from a board, and the value of the questions they receive from the presenter. Watson, without great difficulty, beated the competitors of the program, including samples, and has largely responded in a correct way, interpreting the question directly ruled by the presenter.

Watson was therefore capable of semantic interaction with the presenter and the environment, being able to interpret the meaning of the questions correct? So it seems, but only to a superficial observer. IBM provides in a video and a web site all relevant information to understand the mode of operation of the system. So how does Watson work?

*«A question is analyzed and, based on search in various set of data, a large set of possible answers is extracted. These 'candidate' answers are analyzed separately along a whole series of different dimensions (geographical or temporal dimensions, or, which I found the most interesting, putting back candidate answers into the original question and search that again against various sources of information to rank them again). The result is a vector of numerical values representing the results of the analysis along those different dimensions. That "vector" is summed up into one final value using a weight values for each dimension. The weights themselves are obtained through a prior training process (in this case using a number of stored Jeopardy question/answers). Finally, the answer with the highest value is returned.»[5]*

IBM Watson is an experiment really interesting, and certainly a fabulous step in the direction of the use of the semantic web. Watson responds correctly in almost all occasions, but the positive results, through vastly superior to the capabilities of human competitors, are due to the context of the test. To get a response from Watson is necessary to wait an average of three or four seconds, considerably undue to be transferred to general applications, such as a web search engine, which could not afford to wait so that the user interrogates him. It is said, therefore, that the system developed by IBM, although absolutely remarkable, is not yet the right solution for other environments than those for which it was created, quizzes with prizes.

---

[5] How Watson Works - semanticweb.com - April 24, 2012

**A Semantic Facebook ?**

A recent implementation of Facebook is trying to bring in the same semantic direction the giant social network.

On January 2013, Facebook introduces new status updates on users profile, to help them share what they are feeling, watching, eating, reading, and more. When users go to create a post, Facebook asks «What are you doing?» and includes a drop down menu of options, such as *feeling*, *watching*, *eating* and more, which are then appended to the status update, along with an emoticon, or with a link to the user's page. This feature permits the users to share their activity in a way that can be later used for ad targeting or indexed web search.

In 2011, Facebook introduced the concept of Open Graph applications, for which developers could create custom verbs and publish structured stories about what users did in their apps. This helps Facebook collect important information while giving users new ways to express themselves and learn about things through their friends. Facebook says this information is not being included in Graph Search yet, but it is only natural that it will. We can also envision Facebook using these structured status updates in new types of feeds that focus on a particular category. We've seen the social network adding features to its music dashboard and have wondered whether similar products could be made for movies, books or news. There are other interesting displays that could be created, such as a look at how your friends have been feeling lately. Based on these sentiments and a user's Likes and actions, Facebook could recommend whom to buy gifts for and what those users might want.

**Google's Semantic Search: state of the arts**

At present, the search engine Google, through its voice interface, Google Voice Search works in a syntactic way, responding to the user's query with a result based on a range of probabilities under which a constantly evolving algorithm works. The first intervention of

Google-oriented research that uses semantic criteria was the launch, 16 May 2012, the instrument Knowledge Graph. In the article that explains how it works, Google says: "now Google understands the difference, and can narrow your search results just to the one you mean—just click on one of the links to see that particular slice of results".

Unlike the Google search syntax, the classic one we're used to, which compares the words of a query object with the huge computerized archive, Knowledge Graph analyzes the relationships between words or groups of them, and decide what will be proposed in response to the user. The scheme of operation involves a series of semantic links that would lead the system to understand the real needs of the user connected.

Knowledge Graph is probably one of the largest databases of concepts and relationships however, unlike the aforementioned Watson, the same relationships are not currently exploited to perform a semantic analysis of the query. All it does with the Google Knowledge Graph, at the moment, is to offer the user a simple navigation of its contents. Under some specific query engine extracts information from the database and shows the user the right side of the SERPs (Search Engine Result Page). The user is led to believe that the relative box on the right are selected through a semantic analysis, while it was simply made a statistical analysis of the giant database query, to determine which of them deserved the box deepening and what types of information, for each of them, it was appropriate to present in the box.

"With the Knowledge Graph, Google can better understand your query, so we can summarize relevant content around that topic, including key facts you are likely to need for that particular thing. For example, if you are looking for Marie Curie, you'll see when she was born and died, but you'll also get details on her education and scientific discoveries. How do we know which facts are most likely to be needed for each item? For that, we go back to our users and study in aggregate what they've been asking Google about each item. For example, people are interested in knowing what books Charles Dickens wrote, whereas they're less interested in what books Frank Lloyd Wright wrote, and more in what buildings he designed. The Knowledge Graph also helps us understand the relationships between things. Marie Curie is a person in the Knowledge Graph, and she had two children, one of whom also won a Nobel Prize, as well as a husband, Pierre Curie, who claimed a third Nobel Prize for the family. All of these are linked in our graph. It's not just a catalog of objects; it also models all these inter-relationships. It's the intelligence between these different entities that the key."[6]

So this would be the way for the development of the semantic web? The answer seems negative, despite the words of Google marketing. Statistical analysis of the database query has determined that the writer Charles Dickens was appointed by the users often specifying the object of the requested information, for example what books he wrote. This allows Google to know 'books' is a term in the query is frequently associated with Charles Dickens and so enough for him to establish that an answer to the query [Charles Dickens] could benefit from the list of the works of the writer.

Google therefore has a large archive of concepts and relationships, but does not use it for the stated goal, namely the semantic analysis neither to propose user the semantic results of the research.

**Pragmatic web and applications: limited results**

After this quick analysis of the tools they would use a semantic approach in the interaction with the user, or at least that are announced as such, we check how it is used in the web and pragmatic with any electronic instruments.

The phrase "My daughter is a fox", determines a number of Google results for about 500,000 pages. One would expect to find results related to cunning, guile, speed of reasoning, if not of a daughter, at least of a human being.

On the contrary, however, the main results suggest pages where we discuss foxes, people who call themselves "Fox", and experience with the daughter of someone with a copy of the canid.

The most obvious conclusion, after some similar experiments on the same search engine, is that in reality we are extremely far from any use of pragmatics by machines.

From the pragmatic side, the machines can not '*understand*' the inferences, and, despite the attempt to connect words and sentences with meta data, fail to complete those simple arguments that allow us to fully understand an utterance as "my daughter is a fox". The machines fail to gather information for a pragmatic approach, which instead plays a human being to come to a full understanding of the statement. Among these, for example, knowledge from other statements, the situation of utterance, the intentions of the interlocutor, the specific context refers to the world in general.

---

[6] Singhal, Amit (May 16, 2012). "Introducing the Knowledge Graph: Things, Not Strings". Official Blog (of Google). Retrieved May 18, 2012.

Evi Technologies Ltd. has developed a platform, on the web, *Evi* (www.evi.com), which works by using certain algorithms of a pragmatic nature, to answer questions written by the user. Trying to write "where is Elvis?" Evi says, "Elvis Presley died on August 16th 1977 and is buried in Graceland". To get an answer like, Evi uses an assumption of pragmatic, linking the name with the popularity of appointed. Evi provides a plausible answer, but of course the risk that the user refers to other less well-known Elvis, and thus to provide an answer totally out of place. Evi can provide the reasoning used, with every single step, to give the answer.

**Conclusions**

We analyzed all the newest electronic and computer devices, included newest Google Voice Search and Siri by Apple, many of which are advertised and offered to the user as with semantic capabilities. From what we've seen, however, the semantic field for the machines is still a rather distant goal. Access to information about the world is now easy and affordable, and the machines have access to an immensity of content within which select responses to user requests. The semantics, however, continues to be a difficult goal to achieve, and the cause is to be found in its pragmatic dimension. Despite the enormous progress in understanding the world, the variety of intentions and situations related speech act make it extremely complex relationship between machines and humans, especially when the latter use the conversation context as an integral part of the speech.